\documentclass{article}
\usepackage{graphicx}
\graphicspath{ {./images/} }
\usepackage{comment}
\usepackage{todonotes}
\newcounter{todocounter}

 \title{Using off-the-shelf LLMs to query enterprise data by progressively revealing ontologies}
 
\author{C. Civili (RelationalAI), E. Sherkhonov (RelationalAI), and \\R.~E.~K.~Stirewalt (RelationalAI)}
\date{}

\begin{document}

\maketitle

\section{Introduction}

Using Large Language Models (LLMs) to generate database queries is an area of active research.
In~\cite{sequeda2023benchmark},
Sequeda \emph{et al.}~argue that knowledge graphs (KGs) with rich ontologies can
enable an LLM to answer queries of \emph{enterprise complexity}, noting that text-to-SQL benchmarks such as Spider \cite{yu2018spider} are not tailored to such queries.
In addition to query complexity, an equally challenging problem in the
enterprise setting is
\emph{schema complexity}, where the ontology itself is large and complex.
This paper contributes an approach to using off-the-shelf LLMs and enterprise-scale ontologies to answer natural language questions on large data sets.
We address the schema complexity problem by incrementally revealing "just enough" of an ontology that is needed to answer a given question.
Our approach avoids the need for fine-tuning a model, which may not always be practical, while addressing the problem of LLM hallucinations.

In their paper~\cite{sequeda2023benchmark},
Sequeda \emph{at al.}~recognize the challenges of schema complexity, but this is not reflected in their proposed benchmark, whose ontology is too small and thus not representative of enterprise ontologies.
One challenge, briefly mentioned, is that a much larger ontology would fail to fit the typical context window of an interaction with an LLM.
We observed another challenge -- that the larger and more complex the ontology, the more likely an LLM is to hallucinate.
We believe that addressing these challenges is crucial in building an effective solution for querying data at enterprise scale and we specifically address the problem of fitting a realistic enterprise ontology in the context window of an off-the-shelf LLM.

The intuition behind our approach is that enhancing the prompt with the whole ontology is not only impossible in many realistic cases, but also potentially harmful. What the LLM needs to know is just enough of the ontology to properly translate a natural language query into a formal query. The LLM can help in figuring out exactly which part of the ontology needs to be used to answer the query. This is done via an iterative process that progressively reveals more parts of the original ontology, until the LLM has enough information to produce a formal translation of the query.
Consistent with \cite{sequeda2023benchmark}, we use SPARQL rather than SQL as the formal language for querying our enterprise knowledge graphs.
SPARQL is sufficiently covered in the text corpora used to train LLMs, and it nicely leverages ontologies, which can be verbalized easily using natural language.

We validated our approach on the industrial enterprise KG of a major telecom provider in the US.
Their ontology is on the order of 500 concepts and 1000 relationships \cite{snowflake2023}.
Expressed in OWL, this amounts to more than seven thousand axioms.
Their enterprise KG weaves tens of terabytes of data into that ontology.
We tested queries of increasing complexity, all of which involve navigating long paths and using aggregations.

\section{Related Work}

There are several approaches to leveraging ontologies when using LLMs \cite{pan2024unifying}, two of which are the most prominent. 
One is to fine-tune the model, i.e., to enhance it with the knowledge embedded in the ontology, which typically takes time and is costly.
Another is the zero-shot prompting approach, in which everything the LLM may need to do a task over and above its original knowledge is provided as context to the question itself.
We focus on the latter, as did Sequeda \emph{et al.}~\cite{sequeda2023benchmark}.
Their work shows that using a KG with an ontology in conjunction with an LLM strongly improves the accuracy of natural language querying over an enterprise data set from 16\% (achieved without the use of an ontology \cite{yu2018spider}) to 54\%.
However, the ontology used in their benchmark is very small (11 concepts and less than 30 relationships).
By contrast, enterprise ontologies are typically on the order of hundreds of concepts and thousands of relationships \cite{snowflake2023}.

A realistic medium-size enterprise ontology most likely will not fit the token limitation of an LLM, leaving open the problem of how to share the knowledge embedded into the ontology with the LLM in a single interaction. This problem is briefly mentioned in \cite{sequeda2023benchmark}, where it is said that one would expect to have a RAG (Retrieval Augmented Generation) \cite{lewis2020retrieval} approach to extract the parts of the schema/ontology that are needed to then be provided into the prompt. However, to our knowledge, there is no validation of this solution.

Our approach is similar to RAG, in the sense that we have a retrieval and an enhancement phase, but it has two crucial differences: we use the LLM, as it is, to retrieve the relevant additional knowledge, and we generate on the fly a formalization of such knowledge that is useful for the task . No vector databases are involved in our approach, while RAG usually depends on the capability to successfully encode and retrieve the additional knowledge as vectors. 

Qirk \cite{scheerer2024qirkquestionansweringintermediate} aims to answer natural language questions against a KG using an intermediate representation but without reasoning about its ontology.
Experiments to date with Qirk have focused on simpler queries than those that are expressible in SPARQL. 
Our experience with the telecom provider, in addition to that of~\cite{sequeda2023benchmark}, indicates that features like aggregation and disjunction are commonly needed when answering real user queries over a knowledge graph.

\section{Methodology}

\begin{figure}[h!]
  \includegraphics[width=\textwidth]{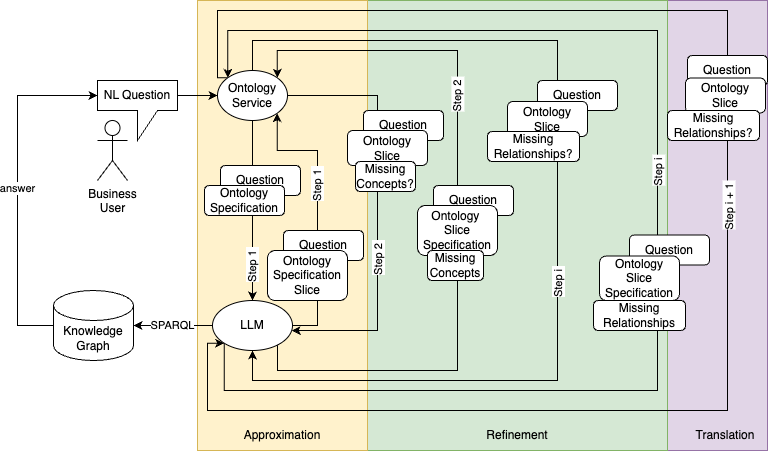}
  \caption{Natural language querying pipeline}
  \label{fig:diagram}
\end{figure}

The key idea underlying our approach is that an LLM should be well-equipped to associate natural language queries to natural language descriptions of an ontology and to associate formal queries to formal descriptions of an ontology.
A second insight is that by reasoning about an ontology while answering a natural language query, we can help an LLM make connections that are valid in the data but that it would not have known about unless it had been
fine-tuned using the ontology.

Building on these intuitions, we split the interaction with the LLM into three phases -- an \emph{approximation phase}, a  \emph{refinement phase} and a  \emph{translation phase}. 
During these phases the LLM works collaboratively with an external service and progressively reveal bigger parts of the ontology, until the resulting slice is sufficiently detailed to answer the original query. 

In the \emph{approximation phase} the LLM is required to complete a single natural language task.
This consists of associating the names of concepts and relationships that are extracted from the natural language description of the ontology to the input question (see Step 1 in Figure \ref{fig:diagram}). 
Notice that an informal description of the ontology typically requires much less space than a formal one, thus the context window limitation is not a concern in this step.
This phase helps in figuring out approximately the relevant part of the ontology needed to answer the query. 

The \emph{refinement phase} consists of one or more steps, each of which requires the LLM to complete a formal language task. 
Each refinement step 
uses an \emph{ontology slice} which is derived on demand by an external \emph{ontology service}, using information that flows out of the previous step---i.e., from the approximation phase in the first refinement step and from the $(i-1)$-th refinement step in the $i$-th step.
Each step proceeds by asking
the LLM to translate the original natural language question into SPARQL,
\emph{only if} the ontology slice is detailed enough to answer the question
(see Steps 2 to i in Figure \ref{fig:diagram}).

If the ontology slice is not detailed enough for the LLM to answer the question, it is asked to list what is missing in terms of missing concepts and missing links by which to connect concepts that it knows about.
By reasoning about missing concepts and missing links, the external ontology service can refine the ontology slice presented to the LLM.
This reasoning often requires path-finding algorithms to discover paths through the ontology that connect up concepts whose links the LLM found to be missing.
The refinement phase terminates when the list of missing components is either empty, or remains unchanged for two steps. 
The former case leads to transitioning to the final translation phase, while the latter signals the impossibility to formally translate the original query.

The \emph{translation phase} consists in the last formal language task of the pipeline (see Step i + 1 in Figure \ref{fig:diagram}).
The task is presented exactly as in the refinement phase, but the output of this phase is a SPARQL query conforming to the last ontology slice presented to the LLM, and thus to exactly that subgraph of the whole ontology that is needed to answer the question.
This query can then be run against the knowledge graph to answer the original natural language question.

We tested our approach on the industrial enterprise KG of a major telecom provider in the US. Users of our approach posed OLAP type of questions that when translated to SPARQL typically involve heavy use of disjunction, graph traversal and aggregation. We observed that vast majority of produced SPARQL queries were of good quality with little or no hallucination.

\section{Future Work}

We believe that incrementally revealing an ontology for natural question answering has benefits also for small to medium size ontologies. In such case, despite fitting the full ontology in the prompt context window may not be a concern, the problem of hallucinations still need to be addressed.
Our approach based on including just enough of the ontology in the slice provided to the LLM can help minimize noise and thus reduce the chance of hallucinations. In the future we aim at validating this on ontologies of suitable size.

\bibliographystyle{abbrv}
\bibliography{bibliography}

\end{document}